\documentclass{PoS}

\title{Non-universal, Non-anomalous $U(1)^{\prime}$ in a Model with Anomaly Mediated SUSY Breaking}

\ShortTitle{Non-universal, Non-anomalous $U(1)^{\prime}$ ... Anomaly Mediated SUSY Breaking}

\author{\speaker{Mu-Chun Chen} \\ 
        Department of Physics \& Astronomy, University of California, Irvine, CA92697-4575, U.S.A.\\
        E-mail: \email{muchunc@uci.edu}}

\author{Jinrui Huang\\
        Department of Physics \& Astronomy, University of California, Irvine, CA92697-4575, U.S.A.  \\
        E-mail: \email{jinruih@uci.edu}}

\abstract{
We construct a Minimum Supersymmetry Standard Model expanded by a non-anomalous family (NAF) $U(1)^{\prime}_{\mbox{\tiny NAF}}$ gauge symmetry. All gauge anomalies are cancelled with no additional exotics other than the three right-handed neutrinos. The FI D-terms associated with the $U(1)^{\prime}_{\mbox{\tiny NAF}}$ symmetry lead to additional positive contributions to slepton squared masses. In a RG invariant way, this thus solves the tachyonic slepton mass problem in Anomaly Mediated Supersymmetry Breaking. In addition, the $U(1)^{\prime}_{\mbox{\tiny NAF}}$ symmetry naturally gives rise to the fermion mass hierarchy  and mixing angles, and  determines the mass spectrum of the sparticles. Our model also provides a counter example to the previous claim that the only $U(1)^{\prime}$ that can give rise to realistic fermion mass hierarchy and mixing pattern must be anomalous. 
}

\FullConference{35th International Conference of High Energy Physics - ICHEP2010,\\
		July 22-28, 2010\\
		Paris France}

\begin{document}

\section{Introduction}

Supersymmetry (SUSY) is one of the most appealing candidates as the new physics beyond the standard model (SM).  Among the mediation mechanisms for SUSY breaking, Anomaly Mediated SUSY Breaking (AMSB)~\cite{ref:SUSYAMSB} turns out to be an extremely predictive framework, in which the soft masses for the sparticles are generated by the conformal anomaly. As a result, {\it all} soft masses are determined entirely by the low energy dynamics. The high predictivity also leads to a severe problem in AMSB models as generically the slepton masses are predicted to be tachyonic, because the electroweak gauge groups, $SU(2)_{L}$ and $U(1)_{Y}$, of the MSSM are not asymptotically free. Squarks do not suffer from the same problem as $SU(3)_{c}$ is asymptotically free.

In Ref.~\cite{Chen:2010tf}, we propose a MSSM model combined with a non-anomalous family (NAF) symmetry $U(1)^{\prime}_{\mbox{\tiny NAF}}$ in the presence of three RH neutrino chiral superfields. The Fayet-Iliololous D-terms associated with the $U(1)^{\prime}_{\mbox{\tiny NAF}}$ symmetry lead to additional contributions to the slepton squared masses, rendering all slepton squared masses positive and thus solving the tachyonic slepton mass problem in AMSB in a renormalization group invariant way~\cite{ref:RGEAMSB}. In addition to solving the slepton mass problem, the $U(1)^{\prime}_{\mbox{\tiny NAF}}$ symmetry plays the role of a family symmetry naturally giving rise to fermion masses and mixing angles through the Froggatt-Nielsen mechanism. The anomaly cancellation conditions give rise to constraints on the $U(1)^{\prime}_{\mbox{\tiny NAF}}$ charges of the chiral superfields, more stringent than in the case of an anomalous $U(1)$.  While there exists an earlier claim~\cite{Ibanez:1994ig} that the $U(1)$ symmetry has to be anomalous in order to generate realistic fermion masses and mixing, we note that counter examples to this claim have been found in Ref.~\cite{ref:gauTrmNeuM, ref:SU5U1, ref:TeVU1} in which it is shown that a non-anomalous $U(1)$ symmetry can be a family symmetry giving rise to realistic masses and mixing angles of the SM fermions. 

\section{The Model}

One of the salient features of AMSB is that its prediction for the soft breaking terms are renormalization group (RG) invariant~\cite{ref:RGEAMSB}. In particular, the scalar squared mass term $(m^2)^{i}_{j}$ is given by, 
\begin{eqnarray}
(m^2)_j^i = \frac{1}{2} m_{3/2}^2 \mu \frac{d}{d \mu} \gamma_j^i \; . 
\end{eqnarray}
In the presence of the $U(1)^{\prime}_{\mbox{\tiny NAF}}$, there are additional Fayet-Illiopolous (FI) D-terms contributions to the scalar squared masses.
Including the additional FI-D term contributions to the scalar masses, the new scalar squared masses at the GUT scale can be written as ~\cite{ref:RGEAMSB} 
\begin{eqnarray}
\bar{m}_{Q}^2 & = & m_{Q}^2 + \zeta q_{Q_i} \delta_{j}^{i} \; , \\
\bar{m}_{u^c}^2 & = & m_{u^c}^2 + \zeta q_{u_i} \delta_{j}^{i} \; , \\
\bar{m}_{d^c}^2 & = & m_{d^c}^2 + \zeta q_{d_i} \delta_{j}^{i} \; , \\
\bar{m}_{L}^2 & = & m_{L}^2 + \zeta q_{L_i} \delta_{j}^{i} \; , \\
\bar{m}_{e}^2 & = & m_{e}^2 + \zeta q_{e_i} \delta_{j}^{i} \; ,  \\
\bar{m}_{H_u}^2 & = & m_{H_u}^2 + \zeta q_{H_u} \; , \\
\bar{m}_{H_d}^2 & = & m_{H_d}^2 + \zeta q_{H_d} \; . 
\label{eqn:FIDMass}
\end{eqnarray}
where the notations used in the above equations can be found in Ref.~\cite{Chen:2010tf}. 

In the presence of the $U(1)^{\prime}_{\mbox{\tiny NAF}}$ symmetry, the Yukawa matrices in the superpotential  are the effective Yukawa couplings generated through higher dimensional operators {\it \`a la} the Froggatt-Nielson mechanism,  
\begin{equation}
\label{eqn:genYukawa1}
Y_{ij} \sim \biggl( y_{ij} \frac{  \Phi }{\Lambda} \biggr)^{3|q_i+q_j+q_H|} \; .
\end{equation} 
We search for solutions of charges that satisfy all six anomaly cancellation conditions and give rise to realistic masses and mixing angles for quarks and leptons, including the neutrinos. Moreover, we impose the condition that all sparticle masses must be positive and that the electroweak symmetry is broken. To check the last conditions, we utilize  SoftSUSY3.1.   

\section{Numerical Results}

Below we show a numerical example with the corresponding $U(1)^{\prime}_{\mbox{\tiny NAF}}$ charges of the chiral superfields summarized in Table ~\ref{tbl:u1Charge2}. 
\begin{table}[b!]
\begin{tabular}{c|c||c|c}\hline\hline\
Field & $U(1)^{\prime}_{\mbox{\tiny NAF}}$ charge & Field & $U(1)^{\prime}_{\mbox{\tiny NAF}}$ charge \\ \hline
$L_1$& $q_{L_1} = 3/2$ & $Q_1$ & $q_{Q_1} = 853/450$ \\ \hline
$L_2$& $q_{L_2} = 1/2$ & $Q_2$ & $q_{Q_2} = -1522/225$ \\ \hline
$L_3$& $q_{L_3} = 1/2$ & $Q_3$ & $q_{Q_3} = 908/225$\\ \hline
$e_1^c$& $q_{e_1} = 31228381/1586700$ & $u_1^c$ & $q_{u_1} = -21278009/1586700$ \\ \hline
$e_2^c$& $q_{e_2} = 29641681/1586700$ & $u_2^c$ & $q_{u_2} = -28164287/1586700$\\ \hline   
$e_3^c$& $q_{e_3} = 26468281/1586700$ & $u_3^c$ & $q_{u_3} = -40540547/1586700$\\ \hline   
$N_1$& $q_{N_1} = -31757281/1586700$ & $d_1^c$ & $q_{d_1} = 10200251/528900$\\ \hline
$N_2$& $q_{N_2} = -31757281/1586700$ & $d_2^c$ & $q_{d_2} = 548909/21156$\\ \hline   
$N_3$& $q_{N_3} = -31757281/1586700$ & $d_3^c$ & $q_{d_3} = 1390561/105780$\\ \hline  
$H_u$& $q_{H_u} = 34137331/1586700$ & $\Phi$ & $q_{\Phi} = -1/3$\\ \hline 
$H_d$& $q_{H_d} = -25674931/1586700$  & $\Psi$ & $q_{\Psi} = 28583881/793350$\\ \hline \hline
\end{tabular}
\caption{The $U(1)^{\prime}_{\mbox{\tiny NAF}}$ charges of the chiral superfields. Note that even though some of the charges for the field $f$ may appear to be very large $\sim \mathcal{O}(20)$, we have the freedom of choosing an overall gauge coupling constant $g$ to be on the order of $< \mathcal{O}(0.1)$ so that the corresponding gauge coupling of the field $f$, $g_{f} = g \cdot q_{f}$, remains perturbative.}   
\label{tbl:u1Charge2}
\end{table}
With these charges, only the diagonal terms in the effective up-type and down-type quark Yukawa matrices are allowed, 
$Y_{u} \sim \mbox{diag}(\lambda^{10}, \lambda^{3}, \lambda^{0})$ and 
$Y_{d} \sim \mbox{diag}(\lambda^{5}, \lambda^{3}, \lambda)$, 
which give rise to the quark mass hierarchy naturally taking into account the $\mathcal{O}(1)$ coefficients. The resulting CKM matrix is an identity, which is to the leading order a good approximation.

To obtain the sparticle mass spectrum, we choose $\zeta = 1.5 \times (100 \; \mbox{GeV})^2$, $\tan \beta = 10$ and $\mbox{sign}(\mu) = -1$ and $m_{3/2} = 40$ TeV, without including the CKM mixing in the quark sector.  Taking the scalar masses shown in Eq. (\ref{eqn:FIDMass}) as the boundary conditions at the GUT scale, we then run SoftSUSY 3.1 and obtain the sparticle masses at the SUSY breaking scale. The sparticle mass spectrum is summarized in Table~\ref{tbl:mass2}.

\begin{table}[t!]
{\small \begin{tabular}{c|c|c|c|c|c|c|c|c|c|c|c|c|c|c|c}\hline\hline\
 $h_0$ & $H_0$ & $A_0$ & $H^+$ & $\tilde{g}$ & $\chi_1$ & $\chi_2$ & $\chi_3$ & $\chi_4$ & $\chi_1^{\pm}$ & $\chi_2^{\pm}$ & $\tilde{u}_L$ & $\tilde{u}_R$ & $\tilde{d}_L$ & $\tilde{d}_R$ & $\tilde{c}_L$ \\ \hline
 115 & 276 & 276 & 287 & 880 & 134 & 362 & 518 & 526 & 134 & 525 & 826 & 795 & 829 & 964 & 743 \\ \hline \hline 
 $\tilde{c}_R$ & $\tilde{s}_L$ & $\tilde{s}_R$ & $\tilde{t}_1$ & $\tilde{t}_2$ & $\tilde{b}_1$ & $\tilde{b}_2$ & $\tilde{e}_L$ & $\tilde{e}_R$ & $\tilde{\mu}_L$ & $\tilde{\mu}_R$ & $\tilde{\tau}_1$ & $\tilde{\tau}_2$ & $\tilde{\nu}_{e_L}$ & $\tilde{\nu}_{{\mu}_L}$ & $\tilde{\nu}_{{\tau}_L}$ \\ \hline
 753 & 747 & 1014 & 367 & 781 & 745 & 905 & 322 & 251 & 298 & 219 & 120 & 299 & 312 & 287 & 286 \\ \hline \hline
\end{tabular}}
\caption{The mass spectrum of the sparticles. All masses are shown in units of GeV.}  
\label{tbl:mass2}
\end{table}

\section{Conclusion}

We propose a MSSM model expanded by a non-universal, non-anomalous $U(1)^{\prime}_{\mbox{\tiny NAF}}$ symmetry. All anomaly cancellations conditions are satisfied with no exotics other than the three right-handed neutrinos. The $U(1)^{\prime}_{\mbox{\tiny NAF}}$ symmetry plays the role of the family symmetry, giving rise to realistic masses and mixing angles for all SM fermions. Furthermore, the FI-D terms associated with the $U(1)^{\prime}_{\mbox{\tiny NAF}}$ symmetry give rise to additional contributions to the slepton masses, rendering them all positive. In a RG invariant way, this thus solves the slepton mass problem in AMSB models. The anomaly cancellation conditions give rise to very stringent constraints on the $U(1)^{\prime}_{\mbox{\tiny NAF}}$ charges of the chiral superfields. We found charges that satisfy all anomaly cancellation conditions and fermion mass and mixing angles, and at the same time solving the slepton mass problem.  While these rational charges are rather complicated, mainly because of the $[U(1)^{\prime}_{\mbox{\tiny NAF}}]^{3}$ anomaly cancellation condition, the differences among the charges are quite simple. The $U(1)^{\prime}_{\mbox{\tiny NAF}}$ charges also dictate the mass spectrum of the sparticles. 

\section*{Acknowledgement}
The work was supported, in part, by the National Science Foundation under Grant No. PHY-0709742 and PHY-0970173.

\end{document}